\documentclass[twocolumn,aps,pre,amssymb,amsfonts]{revtex4-1}
\usepackage{amsmath}
\usepackage{amssymb}
\usepackage{bm}
\usepackage{graphicx}
\usepackage{color}

\begin{document}

\title{Anomalous transport in cellular flows: The role of initial conditions and aging.}

\author{Patrick P\"oschke}
\affiliation{Institute of Physics, Humboldt University of Berlin, D-12489 Berlin, Germany}
\author{Igor M. Sokolov}
\affiliation{Institute of Physics, Humboldt University of Berlin, D-12489 Berlin, Germany}
\author{Alexander A. Nepomnyashchy}
\affiliation{Department of Mathematics, Technion - Israel Institute of Technology, Haifa 32000, Israel}
\author{Michael A. Zaks}
\affiliation{Institute of Physics and Astronomy, University of Potsdam, D-14469 Potsdam, Germany}

\begin{abstract}
\noindent
We consider the diffusion-advection problem in two simple cellular flow models (often invoked as examples for subdiffusive tracer's motion) and concentrate on the intermediate time range, in which the tracer's motion indeed may show subdiffusion. We have performed extensive numerical simulations of the systems under different initial conditions, and show that the pure intermediate-time subdiffusion regime is only evident when the particles start at the border between different cells, i.e. at the separatrix, and is less pronounced or absent for other initial conditions. The motion moreover shows quite peculiar aging properties which are also mirrored in the behavior of the time-averaged mean squared displacement for single trajectories. This kind of behavior is due to the complex motion of tracers trapped inside the cell, and is absent in classical models based on continuous time random walks (CTRW) with no dynamics in the trapped state. 
\end{abstract}

\maketitle

\section{Introduction}

Anomalous diffusion in cellular flows has drawn considerable attention as a model showing a non-trivial 
intermediate asymptotic regime corresponding to subdiffusion, and is reviewed e.g. in \cite{Bouchaud, Isichenko}.  
A simple explanation (e.g. given in \cite{Isichenko}) puts this model into a class of close relatives of the 
comb models. In a comb, diffusion on a spine (say in $x$-direction) is interrupted by the diffusion in the teeth 
(dangling ends) extending in $y$-direction, and can be described by the continuous time random walk (CTRW) in the $x$-direction, with the waiting
times distributed according to a power law. Such diffusion is anomalous (subdiffusion), and, like diffusion in other 
CTRW models with power-law waiting times, shows aging and weak ergodicity breaking, 
see e.g. \cite{Meroz}. Moreover, the properties of such diffusion strongly depend on the initial conditions.  

The similarity between the cellular flow and the comb structures arises from the fact that being trapped inside a
single eddy cell, the particles cannot take part in the macroscopic diffusion unless returning to the separatrix between the cells, 
so that the coarse-grained picture of the macroscopic motion is again the CTRW. In this respect, diffusion in 
a cellular flow corresponds to the comb model with finite teeth, since the power-law waiting time on a site (inside the cell)
has an exponential cutoff implied by the finite size of the trapping domain. Under such conditions the final
asymptotic behavior of transport should be diffusive, with the effective diffusion coefficient 
$D^* \propto D \mathrm{Pe}^{1/2}$, where $D$ is the coefficient of molecular diffusion, and $\mathrm{Pe}$ is the P\'{e}clet number, \cite{Pomeau,Childress,Soward,Shraiman,RosenbluthEtAl}.
The employed mathematical approaches are based on homogenization (see e.g. \cite{Majda} and further references therein) and on the large deviation theory \cite{HaynesVanneste2014}. 
Analytical predictions are compared to numerical simulations of the transport in the corresponding time range.

Remarkably, the prominent subdiffusive intermediate asymptotics stays much less investigated. The diffusion in cellular flows in such 
regime is sometimes considered as an example for a situation where aging and ergodicity breaking
take place, see e.g. \cite{Akimoto1,Akimoto2}, without paying much attention to the peculiarity of this problem
caused by the complicated dynamics inside the trapped state. 
We are not aware of extensive numerical simulations of the stochastic differential equations for these systems and the comparison of the numerics with the theoretical predictions of Ref. \cite{Pomeau}
and subsequent works even for the simplest initial conditions corresponding to starting on the cell boundary
(separatrix), as well as of any work considering aging and convergence to ergodic behavior in this context. 
Our work is aiming at filling this gap. In what follows two standard variants of cellular flows are considered,
the eddy lattice (EL) flow in two dimensions, see e.g. \cite{RosenbluthEtAl}, with the streamfunction given by
\begin{equation}
 \psi_{\mathrm{EL}}(x,y) = ua \sin\left( \frac{x}{a}\right)\sin\left( \frac{y}{a}\right)
\end{equation}
and a flow corresponding to a one-dimensional arrangement of cells along the $x$-axis, with no-slip boundary conditions at $y=0$ and $y=\pi a$.
The model streamfunction is
\[
  \psi_{\mathrm{YPP}}(x,y) = ua \sin\left( \frac{x}{a}\right) \left(\frac{y}{\pi a}\right)^2\left(1- \frac{y}{\pi a}\right)^2.
\]
Introduced in \cite{Pomeau},
this flow is referred below to as the Young-Pumir-Pomeau (YPP) flow. In both cases $u$ is the characteristic velocity, and $\pi a$ is the cell size. Both flows show the intermediate-time subdiffusive behavior
crossing over to normal diffusion at longer times. 
The comparison of these two flows is of considerable interest, since the transport behavior in them is very similar in certain respects, but very different in other ones. 

Below, in Sect.~\ref{sect_times} we review the subdiffusion dynamics from the point of view of 
characteristic timescales. The problem is formulated and numerical procedures for direct 
simulation of stochastic differential equations are introduced in Sect.~\ref{sect_sim}.
Results, concerning different aspects of transport, are presented in Sect.~\ref{sect_results}.
Finally, in Sect.~\ref{sect_conclusions} we summarize our findings.

\section{Characteristic times}
\label{sect_times}

Let us first shortly discuss the type of subdiffusive behavior, and the characteristic times at which it can be observed. 
This discussion is a slight modification of the one in Ref. \cite{Pomeau}, so that no more detail as necessary is given here. 
The corresponding results are not new, but will be of importance in what follows. 

The trapping of the particle inside the flow cell is due to its diffusion in the direction transverse to the streamlines,
since, having entered the cell, the particle cannot leave it unless returning to its periphery. The particle motion 
in the direction normal to the streamlines (denoted below as the $z$-direction, with $z=0$ corresponding to the separatrix) is obtained by averaging the diffusion-advection 
equation along the streamlines \cite{Pomeau}. The averaged equation has the form of the diffusion equation
\begin{equation}
 \frac{\partial}{\partial t} p(z,t) = \frac{\partial}{\partial z}\left[ D(z) \frac{\partial}{\partial z} p(z,t) \right].
 \label{diff}
\end{equation}
The $z$-dependence of the diffusion coefficient is sensitive to the boundary conditions on the cell 
edges and therefore depends on the type of the flow. For the free boundaries of the EL flow  $D(z)$ can be taken constant and equal to the coefficient of the molecular diffusion,
for the no-slip boundary conditions on the horizontal sides of the cell of the YPP flow  one has to assume
$D(z) \simeq z^{-1}$ for $z \ll a$, see \cite{Pomeau}.

The (intermediate) asymptotic behavior of the waiting time probability density function (WTD) for the jumps from separatrix to separatrix can be obtained by first solving Eq. (\ref{diff}) for the probability of being at the origin $p(0,t)$ and then connecting this with the first return probability via the renewal approach, assuming the Markovian
character of the process, by solving the integral equation $p(0,t) = \delta(0)\delta(t) + \int_0^t \phi(t') p(0,t-t') dt'$, which can be
easily done in the Laplace domain. The power-law decay of $p(0,t) \propto t^{-\gamma}$ then translates into the behavior 
\begin{equation}
\phi(t) \propto t^{\gamma-2}
\label{WTD}
\end{equation}
for the first return time probability density. Note that this discussion corresponds to the case when the particle starts on the separatrix
($z=0$), and is pertinent to all transitions between the cells. The first waiting time will differ in all cases when the particle does
not start at the separatrix, i.e. for different initial conditions (e.g. starting at the center) or in aged situations \cite{KlaS}.

The behavior of $p(0,t)$ is 
\[
 p(0,t) \propto \left\{ \begin{array}{ll}
  t^{-1/2} & \mbox{for the EL flow} \\
  t^{-1/3} & \mbox{for the YPP flow,}
 \end{array}
 \right. 
\]
which translates into $\phi(t) \propto t^{-3/2}$ (normal Sparre-Andersen behavior) for the EL flow and into $\phi(t) \propto t^{-5/3}$ 
for YPP flow, respectively. 
Considering the corresponding CTRW between the cells with the step size $a$ yields subdiffusion with 
\begin{equation}
 \langle R^2(t) \rangle \propto t^{1-\gamma} = \left\{
 \begin{array}{ll}
  t^{1/2} & \mbox{for the EL flow} \\
  t^{2/3} & \mbox{for the YPP flow.}
 \end{array}
 \right.
 \label{anodiff}
\end{equation}
The power-law behavior as given by Eq. (\ref{WTD}) takes place in a finite time range bounded from below and from above
by two characteristic times $t_1$ and $t_2$. The longer characteristic time $t_2$ corresponds to the time of free molecular diffusion over the
cell length, $t_2 \sim a^2/D$, and for $t>t_2$ the WTD $\phi(t)$ shows an exponential cutoff \cite{Pomeau}. This time marks
the crossover from the anomalous regime to normal diffusion.

The shorter characteristic time $t_1$ does not follow from the pre-averaged Eq. (\ref{diff}), and is the minimal time necessary
to traverse the cell. It is different for the cases of the free and no-slip boundary conditions. 
Note that this minimal time defines the normalization constant of the WTD $\phi(t)$. Since $\phi(t)$ vanishes rapidly both for 
$t \ll t_1$ and for $t \gg t_2$ one has 
\[
 \int_{t_1}^{t_2} \phi(t) dt \simeq 1,
\]
where, due to the fact that $\phi(t) \simeq t^{\gamma-2}$ is integrable, the upper bound 
is irrelevant, provided $t_2 \gg t_1$. Therefore 
\begin{equation}
 \phi(t) \simeq t_1^{1-\gamma} t^{\gamma-2}.
 \label{psi}
\end{equation}
The shorter cutoff time thus defines the coefficient of the anomalous diffusion in the anomalous regime
\begin{equation}
 \langle R^2(t) \rangle \simeq \frac{a^2}{t_1^{1-\gamma}} t^{1-\gamma} 
 \label{anom}
\end{equation}
up to a numerical constant. 
For the free boundary the minimal travel time is defined by the characteristic velocity of the flow and is of the order of $t_1 = a/u$.
For the no-slip condition this is no more the case since the velocity in the boundary region vanishes. 

Let us consider particles in the YPP flow entering a cell on its left side. During the time $t \sim a/u$ of travel along the vertical border of the cell the particle
can move diffusively in the horizontal direction at a distance $\delta \sim \sqrt{Dt} \simeq (Da/u)^{1/2} \propto a \mathrm{Pe}^{-1/2}$, 
with $\mathrm{Pe} = ua/D$ being the P\'{e}clet number of the flow. 
This is the thickness of the boundary layer, as discussed in Appendix C of \cite{Pomeau}. The streamline at such distance from
the vertical boundary is characterized by the value of its streamfunction $\psi \simeq A u \delta$ with $A$ being a number constant. When moving parallel to the horizontal boundary (say, for $x=1/2$, where $\phi = By^2$) a tracer passes at the distance $y \propto \sqrt{a \delta}$ from it. 
The typical velocity at this horizontal part of the streamline is $v_x \simeq u (y/a) \sim u \sqrt{\delta / a}$. Therefore the typical transport time in the horizontal direction would be
\[
 t_1 \simeq \frac{a}{v_x} \propto \frac{a}{u} \mathrm{Pe}^{1/4}.
\]
The ratio of the times $t_2$, which defines the end of the subdiffusion regime and the transition from anomalous diffusion to normal diffusion, and $t_1$,
\[
 \frac{t_2}{t_1} \propto \frac{a^2}{D} \frac{u}{a} \mathrm{Pe}^{-1/4} = \mathrm{Pe}^{3/4}
\]
is a growing function of the P\'{e}clet number, so that there is enough time for developing diffusion anomaly at high P\'{e}clet numbers in a YPP flow. 

Summarizing these findings, we have 
\[
 \frac{t_2}{t_1} \simeq \left\{ \begin{array}{ll}
  \mbox{Pe} & \mbox{for the EL flow} \\
  \mbox{Pe}^{3/4} & \mbox{for the YPP flow.}
 \end{array}
 \right. 
\]
This means that the anomalous diffusion is only pronounced in the case of high P\'{e}clet numbers.

Existence of the upper cutoff guarantees the convergence of the mean waiting time $t^*$ within the cell, given by

\begin{eqnarray*}
 t^* &=& \int_{t_1}^{t_2} t \phi(t) dt \simeq t_1^{1-\gamma} t_2^\gamma \\
 &=& t_2 \left( \frac{t_2}{t_1} \right)^{\gamma-1} = \frac{a^2}{D} \left( \frac{t_2}{t_1} \right)^{\gamma-1}
\end{eqnarray*}

for $t_2 \gg t_1$ and for $0 < \gamma < 1$. The diffusion coefficient in the final regime of normal diffusion is then
$ D^* \simeq a^2/t^* = D \left( \frac{t_2}{t_1} \right)^{1-\gamma}$. Inserting the expression for $t_2/t_1$ and the values of $\gamma$ we arrive at
\[
 D^* \simeq D \mbox{Pe}^{1/2}
\]
for both flows. This form of $D^*$ guarantees the smooth crossover from the anomalous behavior as given by Eq. (\ref{anom}) to the diffusive behavior 
$\langle R^2(t) \rangle \sim D^* t$ at the crossover time $t_2$. 

We note that for the initial conditions different from starting on the separatrix, and for aged situations the distribution of the first waiting time until the jump between the cells $\phi_1(t)$
differs from $\phi(t)$, and cannot in general be obtained within the approach based on the streamline averaging. 

\section{Simulation}
\label{sect_sim}

The particle motion under the influence of the flow and of molecular diffusion is described by the Langevin equation
\[
 \dot{\mathbf{r}} = \mathrm{rot} \; (0,0,\psi(\mathbf{r})) + \sqrt{2D} \bm{\xi}.
\]
Here $\mathbf{r}$ is the instantaneous two-dimensional particle position, and $\bm{\xi} = (\xi_x, \xi_y)$ is a vector of two independent
Gaussian noises with zero mean, unit width and with $\langle \xi_x(t) \xi_x(t')\rangle =  \langle \xi_y(t) \xi_y(t')\rangle = \delta(t'-t)$. 
Taking $a$ as the unit length and $t_2 = a^2/D$ as time unit, we can rewrite this equation as
\begin{equation}
 \dot{\mathbf{r}} = \mathrm{Pe} \; \mathrm{rot} \; (0,0,\Psi(\mathbf{r})) + \sqrt{2} \bm{\xi}
 \label{Lang1}
\end{equation}
with 
\begin{equation}
\Psi(x,y) =\left\{ \begin{array}{ll}
  \sin (x) \sin (y) & \mbox{for EL flow} \\
  \sin (x) \left(\displaystyle\frac{y}{\pi}\right)^2 \left(\displaystyle 1-\frac{y}{\pi}\right)^2 & \mbox {for YPP flow.}
 \end{array} \right.
 \label{Langevin}
\end{equation}
Equation (\ref{Lang1}) is numerically integrated using the Heun algorithm, and the results, the particle trajectories $\mathbf{r}(t)$, are used for further analysis. 
The time step of integration is chosen so, that in the absence of noise the displacement in the direction normal to streamlines is negligible during the full simulation
time $t_{\mathrm{max}}=100$. For both systems taking a time step equal to $1/(1000\mathrm{Pe})$ turned out to be sufficient. 
A typical trajectory for the EL flow is shown in Fig.~1. 

\begin{figure}[t]
\label{Fig_walk}
\centering
\includegraphics[width=80mm]{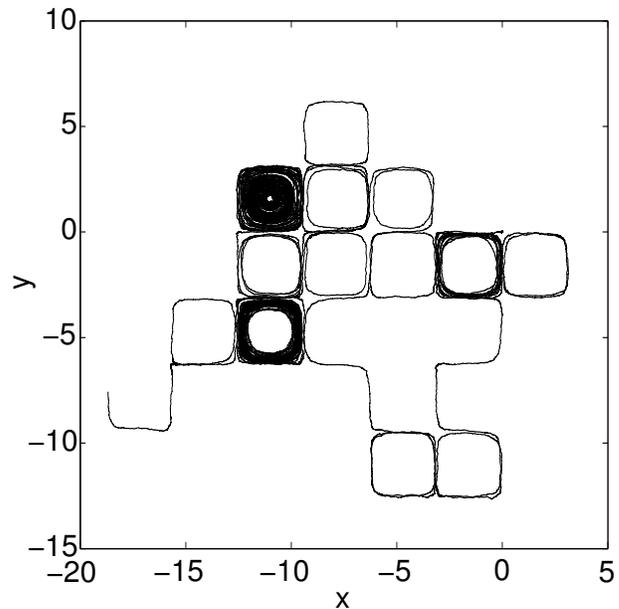}
\caption{A typical walk for the EL flow for Pe $=10^3$}
\end{figure}

The main quantities of interest are the mean squared displacement (MSD) of particles from their initial positions
\begin{equation}
\langle R^2(t) \rangle = \langle (\mathbf{r}(t) - \mathbf{r}(0))^2 \rangle
 \label{MSD}
\end{equation}
for different initial conditions, the aged MSD (the MSD from the 
position a particle had at the time $t_a$ from the beginning of observation)
\begin{equation}
\langle R^2(t,t_a) \rangle = \langle (\mathbf{r}(t_a+t) - \mathbf{r}(t_a))^2 \rangle
 \label{AMSD}
\end{equation}
respectively the root mean squared displacement (RMSD) which is the square root of the MSD,
for $\Delta > 0$, the time-averaged MSD (TAMSD)
\begin{equation}
 \overline{R^2(\Delta, T)} = \frac{1}{T-\Delta}\int_0^{T-\Delta} [\mathbf{r}(t'+\Delta)-\mathbf{r}(t')]^2\, dt',
 \label{TAMSD}
\end{equation}
and its ensemble-averaged analog
\begin{equation}
 \langle \overline{R^2(\Delta, T)} \rangle = \frac{1}{T-\Delta}\int_0^{T-\Delta} \!\!\!\!\!\!\!\!\langle [\mathbf{r}(t'+\Delta)-\mathbf{r}(t')]^2 \rangle dt',
 \label{ATAMSD}
\end{equation}
(for the essentially one-dimensional transport by YPP flow we only consider the displacements along the $x$-axis),
and the probability density function (PDF) of the displacements from the initial position. The ensemble
averaging corresponds to averaging over independent runs. If not stated differently, averaging over $10^4$ independent tracers for Pe $=10^4$ was performed. 

\section{Results}
\label{sect_results}

The results for the main quantities of interest are shown in Figs.~2 to~4 and~6. 
The left panels of these figures present the results for the EL flow, the right ones for the YPP flow. 

\begin{figure*}[]
\label{Fig_MSD}
\centering
\includegraphics[width=80mm]{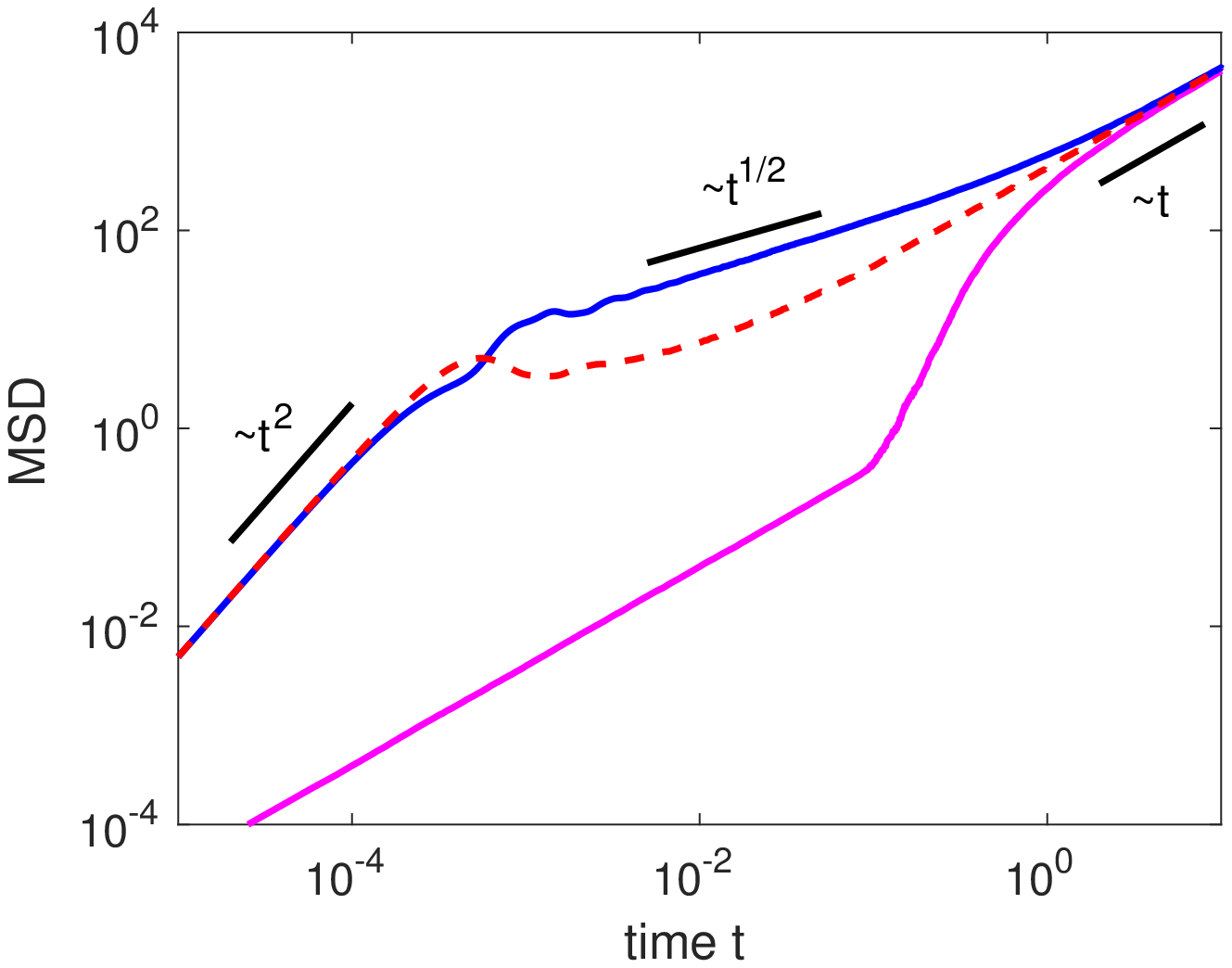}
\includegraphics[width=80mm]{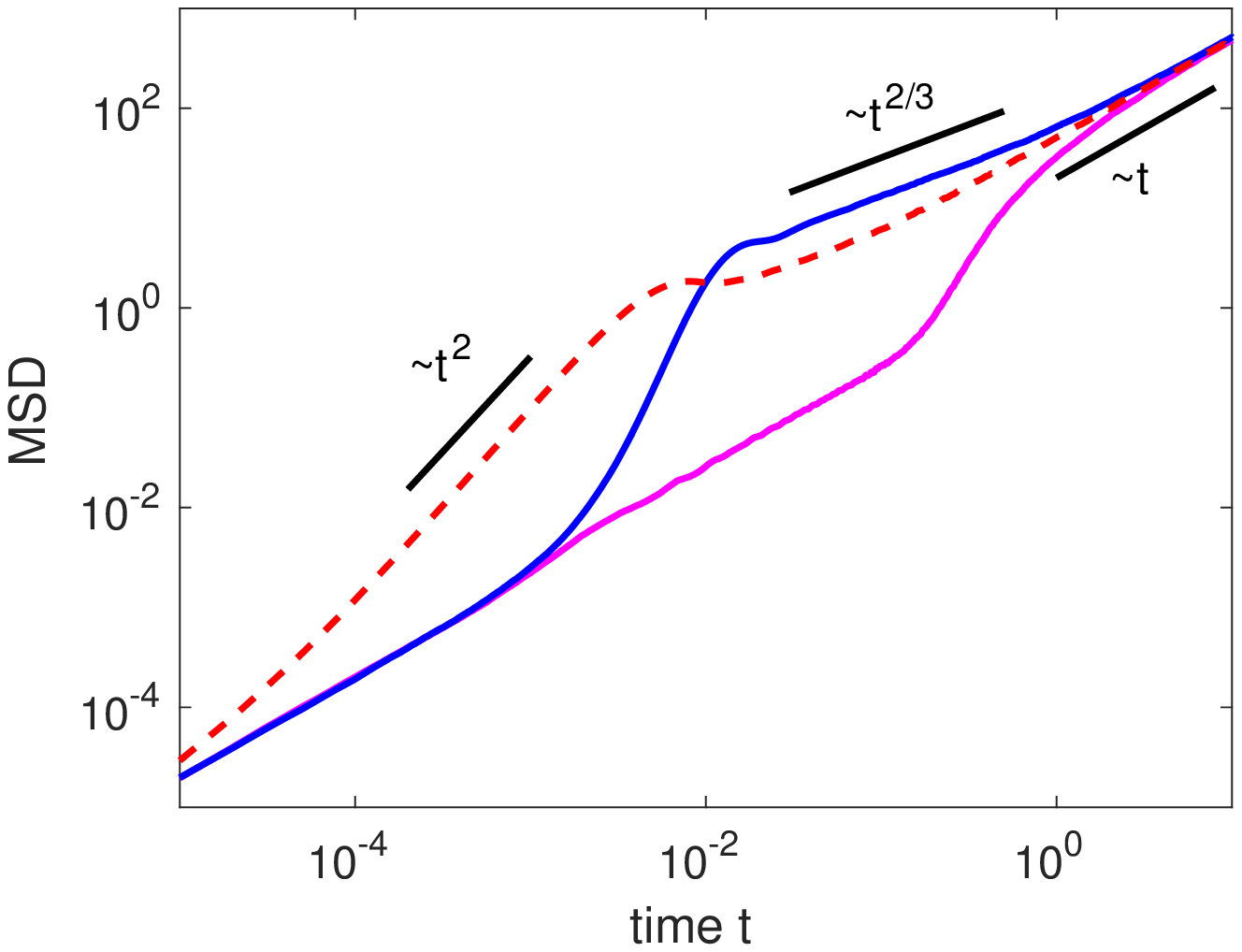}
\caption{(color online) MSD for the EL flow (left) and YPP flow (right) for starting at the separatrix (upper, blue), 
the flooded case (dashed, red), and starting at the center of a cell (lower, magenta).}
\end{figure*}

\begin{figure*}[]
\label{Fig_CDF}
\centering
\includegraphics[width=80mm]{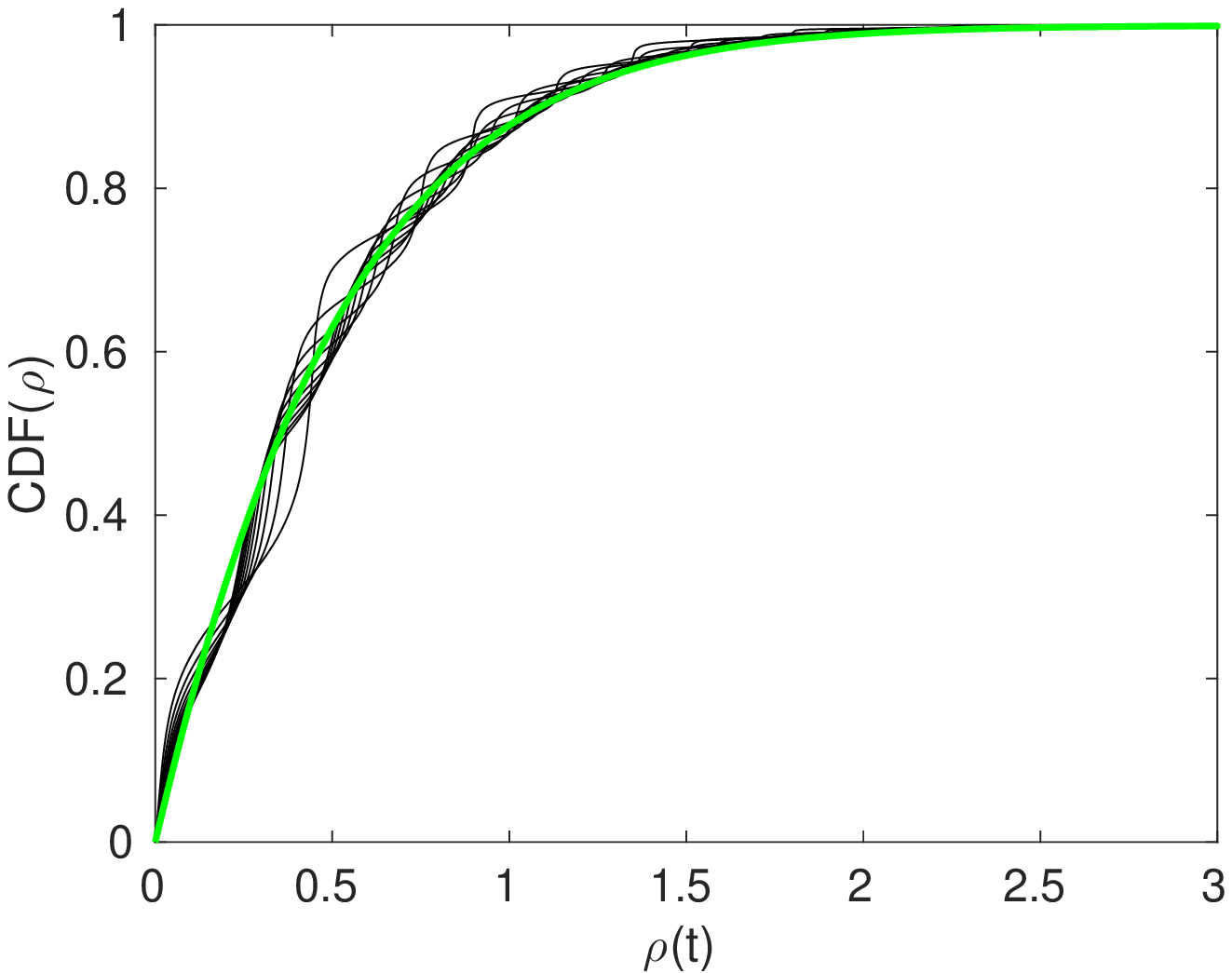}
\includegraphics[width=80mm]{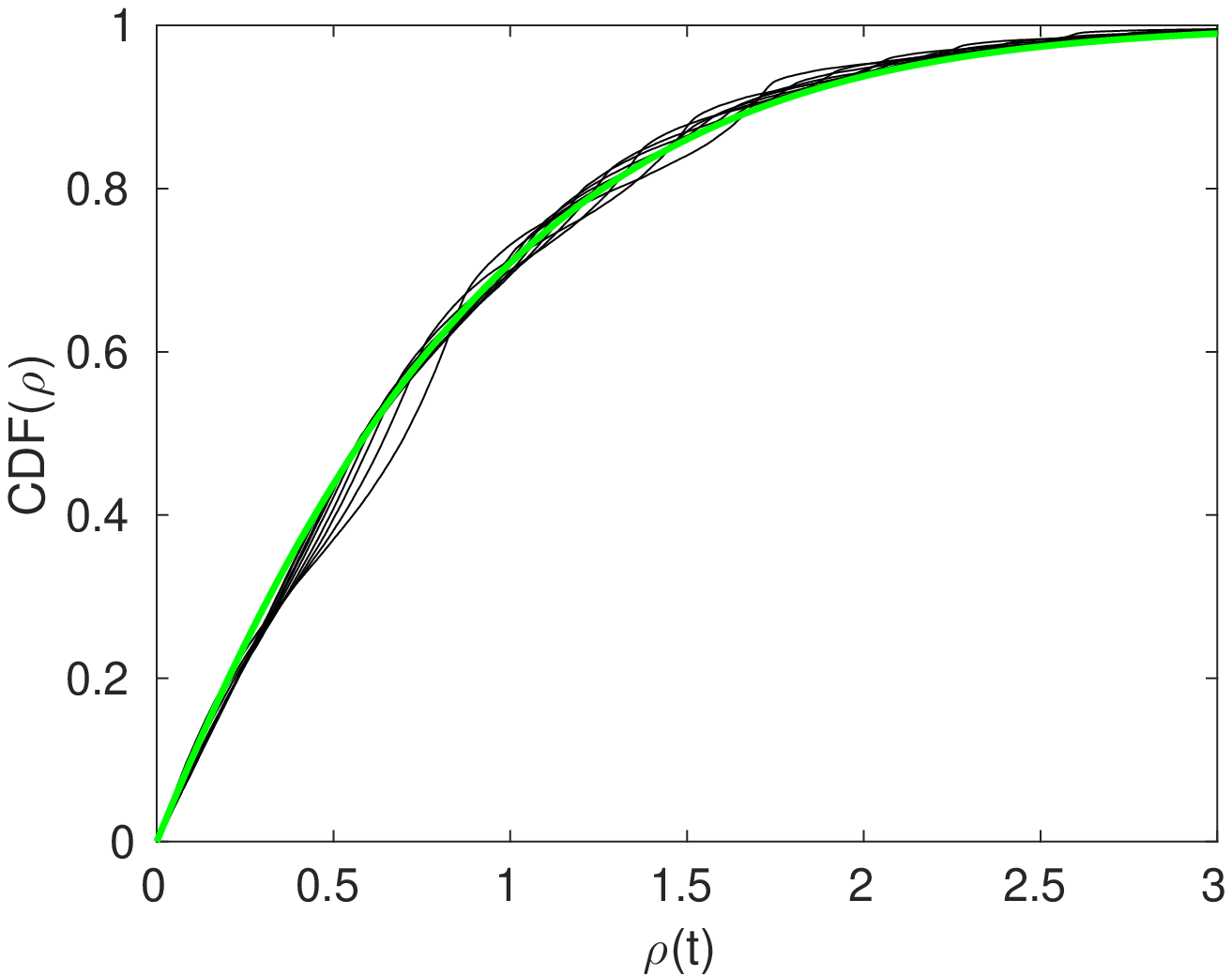}
\caption{(color online) Left panel: CDF depending on the rescaled distance $\rho$ for EL in subdiffusive regime: $10^6$ simulated walks with time $t\in[0.01,0.1]$ in steps of $0.01$ (thin, black) compared to the predictions of 
CTRW (thick, green). Right panel: The same for the YPP flow: $10^5$ simulated walks with time $t\in [0.1, 0.5]$ in steps of $0.05$. 
Note that no adjustable parameters were used to obtain these two plots. }
\end{figure*}

Figure~2 shows the MSD for the corresponding flows for different initial conditions. 
The upper (blue online) curves correspond to the situation where at the beginning the particles were homogeneously
distributed on the separatrix, i.e. have $x(0) = 0$ and $y(0)$ homogeneously distributed between 0 and $\pi$. The lower curves
(magenta online) correspond to starting at the center of the cell, and the dashed curves (red online) correspond
to tracers initially homogeneously distributed within one cell. The latter situation will be referred to as a ``flooded case''.

Let us first discuss the behavior of the MSD for a start at the boundary of the cell, that, for $t > t_1$, is well described by CTRW.
The curves exhibit three distinct regimes for $t < t_1$, for $t_1 < t < t_2$ and for $t > t_2$.
For the EL flow the motion for $t < t_1$ is dominated by the ballistic transport on the cell periphery. 
For the YPP flow the transport at short times (i.e. in the boundary layer) is
dominated by the diffusion and not by the flow, so that the behavior for $t < t_1$ is diffusive. The transition from this domain to the
next one corresponds to a superballistic motion, since the transport velocity grows when the particle 
diffuses into the cell interior.
The following regime of anomalous diffusion corresponds to $\langle R^2(t) \rangle \propto t^{1/2}$ for the EL flow, and to $\langle R^2(t) \rangle \propto t^{2/3}$
for the YPP flow, as predicted by the CTRW model. The behavior for $t > t_2$ is diffusive for both flows. 

The time evolution of the ensemble of tracers starting at the separatrix is indeed well-described by CTRW with the waiting time densities 
$\phi(t) \propto \tau^\alpha t^{-1-\alpha}$ as given by Eq. (\ref{psi}), with the characteristic time $\tau$ being of the order of $t_1$ and $\alpha = 1-\gamma$. 
Given $\phi(t)$ and the mean squared displacement per step $a^2$, the probability density function of the particle displacement in the Fourier-Laplace representation is
\[
 p(k,s) = \frac{\tau^\alpha s^{\alpha-1}}{k^2a^2/2 + \tau^\alpha s^\alpha}
\]
for both $k$ and $s$ small, that corresponds to the long time and large scale limit in the space-time domain, see e.g. Chap. 4 of Ref. \cite{KlaS}. 
The corresponding PDF $p(x,t)$ in space-time domain is an even function of its argument and scales as a function of $\rho = x/R(t)$ where $R(t) = \langle R^2(t)\rangle^{1/2}$
is the root mean squared displacement: $p(x,t) \propto f_\alpha[|x/R(t)|]$, with the scaling function $f_\alpha(\rho)$ depending on the index $\alpha$ as a parameter. 
For the YPP flow this can be found in quadratures \cite{Pomeau}:
\[
f_{2/3}(\rho) \propto \mathrm{Ai}(\rho)
\]
 where $\mathrm{Ai}(\rho)$ is the Airy function, see Chap. 10 in \cite{Abramowitz} (note that $\rho > 0$). For the case of the EL flow ($\alpha = 1/2$) we are aware of no closed form, 
but a useful integral representation (see Chap. 6 of Ref. \cite{KlaS}) helps to
find $p(x,t)$ (and thus $f_{1/2}(\rho)$) numerically: 
\begin{equation}
 p(x,t) \propto \int_0^\infty \frac{1}{\sqrt{K\omega t}}\exp\left(-\frac{x^2}{4K\omega}-\frac{\omega^2}{4t}\right)\, d\omega
 \label{Subord}
\end{equation}
with $K$ being proportional to the coefficient of the anomalous diffusion defining the MSD.  
Knowing $f_\alpha(\rho)$ we can build the corresponding cumulative distribution function (CDF) of the scaled absolute displacements $\rho$, 
\[
 F_{\alpha}(\rho) = \frac{\int_0^\rho f_{\alpha}(z) d z}{\int_0^\infty f_{\alpha}(z) d z},
\]
and compare it to the numerical results, as shown in Fig.~3 
(for the EL flow only the displacement along the $x$-direction is considered). The
corresponding theoretical curves are shown as thick lines (green online). The results of the numerical evaluation of the corresponding CDFs of rescaled distances
at different times are shown with thin lines. These do indeed roughly follow the CTRW predictions, but show additional oscillations, which are not
errors or artefacts, but stem from the internal dynamics of particles within the cells, which is not resolved on the scales where the CTRW approach
is applicable. This is exactly this intracell dynamics which makes the anomalous diffusion in cellular flow different from the one in 
combs with finite teeth.
Note also that the theoretical curves of the CDF$(\rho)$ for different not too small Pe should all coincide as well respectively for both systems. We checked this to hold for ten times larger Pe, i.e. Pe~$=10^5$. \\
\\
Let us return to our discussion of Fig.~2. 
When starting at the center of the cell (lower curves, magenta online), no intermediate subdiffusion is seen, and the behavior 
in both EL and YPP flows corresponds to a superballistic
crossover from the short time diffusion with the diffusion coefficient $D$ to final diffusion with $D^* \simeq D \mathrm{Pe}^{1/2}$.

When the initial positions of the particles are homogeneously distributed within the cell (dashed, red online), the long time behavior is exactly like in the previous two cases. The short time behavior is ballistic, which for the EL flow coincides with the one obtained by a start at the separatrix. Note also that for the flooded case no intermediate domain with a constant diffusion exponent smaller than unity can be detected at all: The crossover from the initial ballistic to final diffusive behavior involves only a
slight oscillation.
A similar behavior of the MSD for the considered initial conditions is seen also for smaller P\'{e}clet numbers (as low as $10^2$), and for the larger ones (up to
$10^6$, the upper limit for our simulations), not shown. \\
\\
For the YPP flow we also take the fourth initial condition of particles starting at a wall, i.e. at $y=0$, see the black dotted lines in the right panel of Fig.~4. 
This initial condition is equivalent to starting at the separatrix with a vanishing advection flow and it corresponds to a pollution model of the atmosphere with dust being initially on earth's surface. The aged MSD for this initial condition is very similar to the one for starting at the separatrix (blue). Only for small times, the MSD is approximately the one for starting at the center of a cell (magenta), because of the initially vanishing advection. \\
\\
The strong dependence on the initial conditions is the reason for aging and for intermediate-time non-ergodic behavior of 
the MSD (see e.g. the discussion in \cite{Sokolov}).

\begin{figure*}[]
\label{Fig_MSDage}
\centering
\includegraphics[width=80mm]{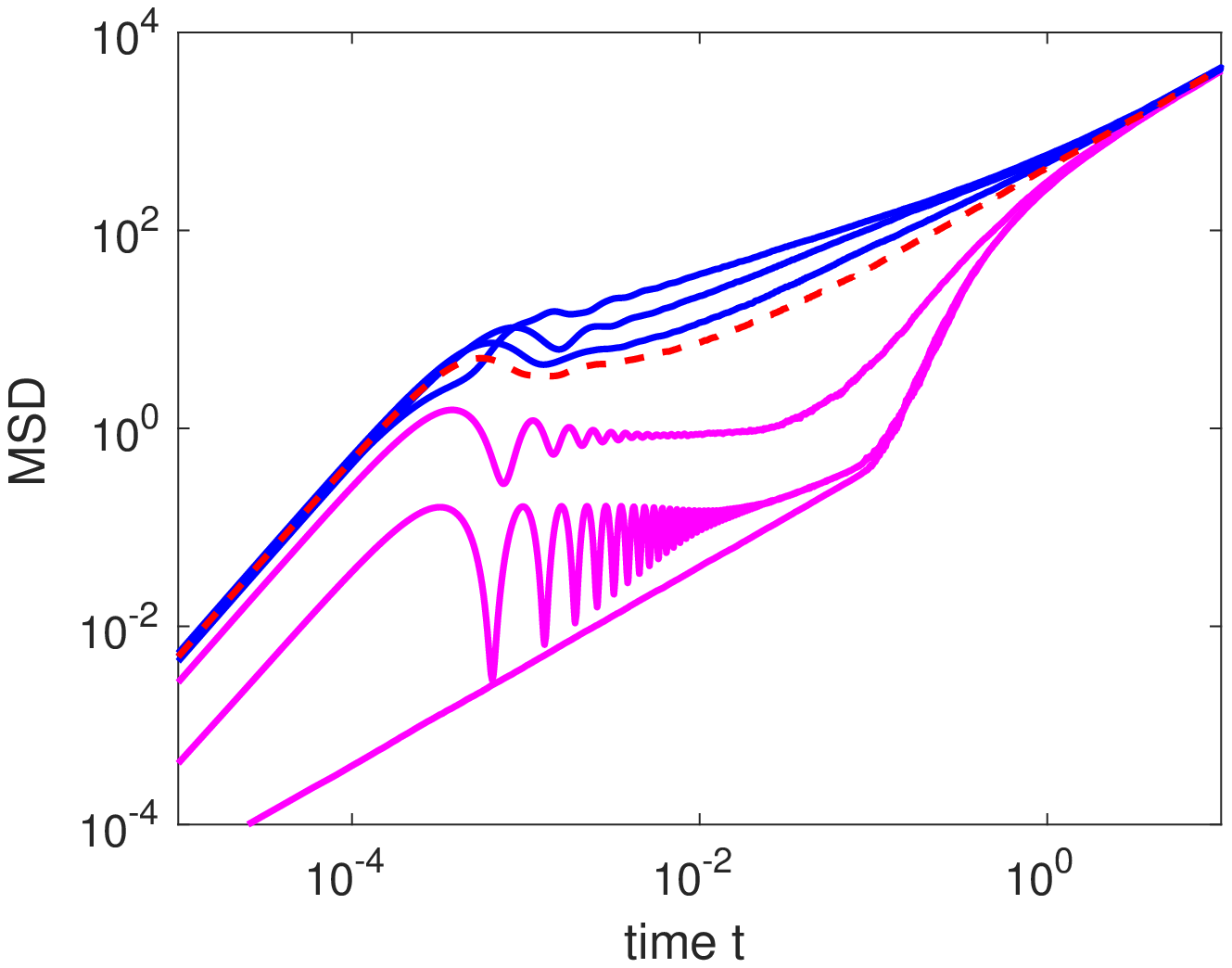}
\includegraphics[width=80mm]{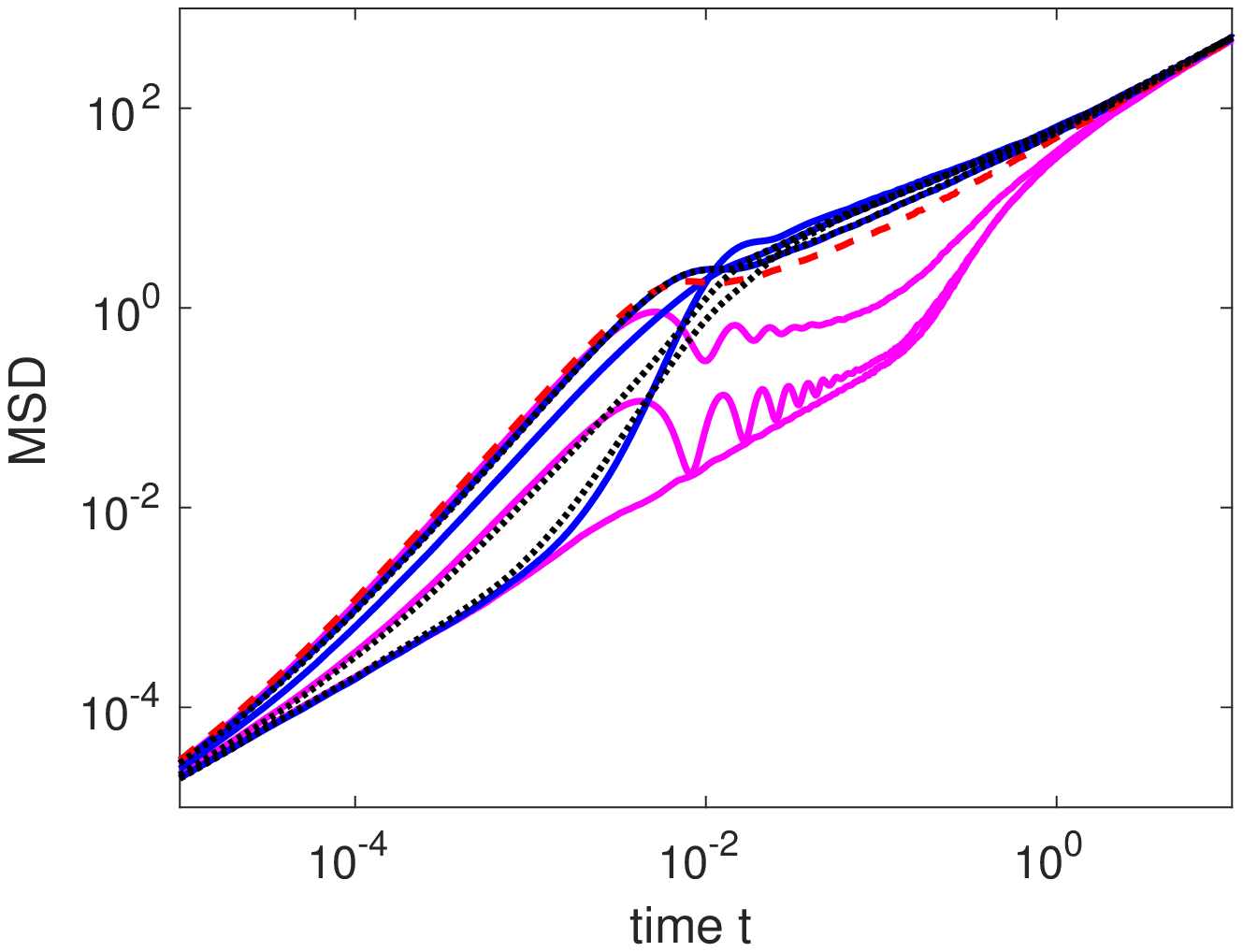}
\caption{(color online) Same as Fig.~2 with aging times $t_a=0$, $10^{-2}$, and $10^{-1}$ (upper, blue and lower, magenta curves) compared to flooded case (dashed, red). In the right panel we show in addition the initial condition of starting at a wall, i.e. at $y=0$ (dotted).}
\end{figure*}

\begin{figure}[t]
\label{Fig_MSDageapprox}
\centering
\includegraphics[width=80mm]{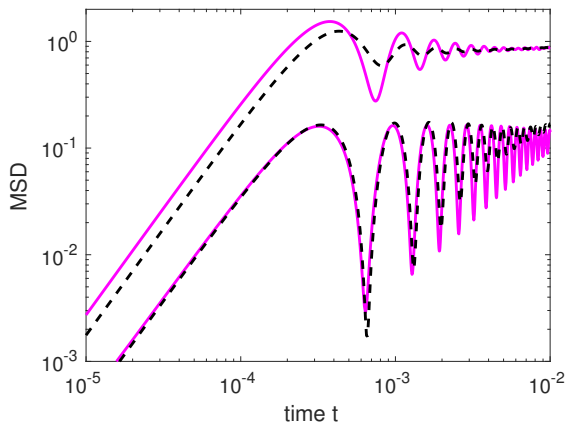}
\caption{Zoom in on the left panel of Fig.~4 compared to approximate expression (dashed).}
\end{figure}

Let us first turn to the aging behavior of the MSD. Since the homogeneous distribution of particles within the system is
invariant under diffusion and flow, the MSD for the initial condition, when the particles are distributed homogeneously
within the cell, does not show any aging effects: $\langle R^2(t,t_a) \rangle = \langle R^2(t) \rangle$. For any other initial condition 
aging is present, as shown in Fig.~4
, and the  $\langle R^2(t) \rangle$ for the homogeneously flooded cell acts as the limiting curve for
$\langle R^2(t, t_a) \rangle$ for $t_a \to \infty$. When starting at the separatrix, this limiting curve (in the intermediate time domain corresponding to
anomalous diffusion) is approached from above for both flows. In the short time domain there is a difference between the EL and the YPP flows
caused by different relative positions of the MSD curves discussed above. Since for the EL flow the short-time behavior for starting
at the separatrix, and for starting homogeneously within the cell coincide, the MSD at short times does not age. On the contrary, for the YPP flow
it shows a considerable speed-up. In the situation when the particle starts at the center of the cell, considerable aging
effects are always observed. For EL the short time behavior of the aged MSD is always ballistic, and approaches the 
asymptotic (ballistic) short-time behavior from below. For the YPP flow the aged short-time MSD behavior shows the change of regime, from
diffusive to ballistic. 

\begin{figure*}[]
\label{Fig_EATAMSD}
\centering
\includegraphics[width=80mm]{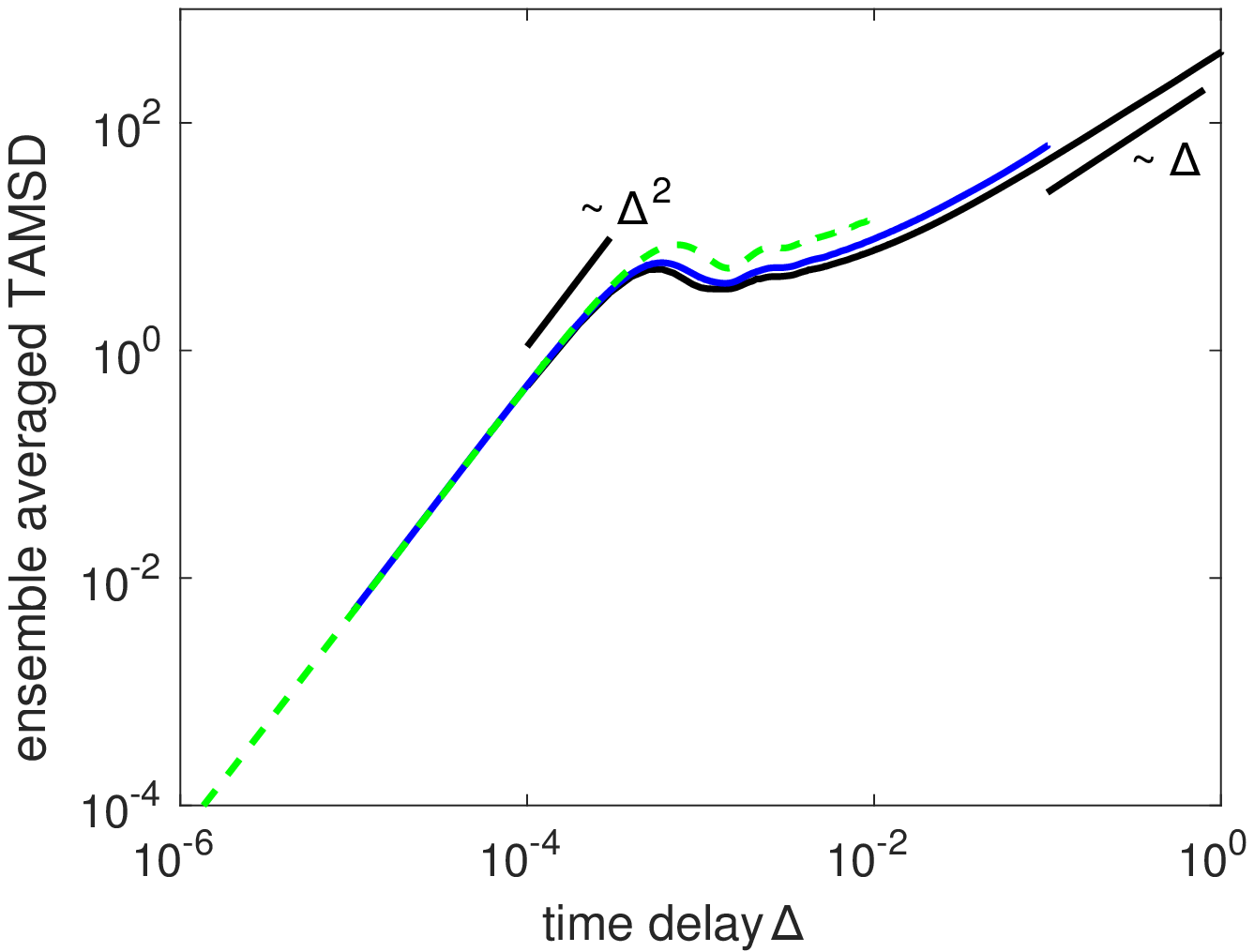}
\includegraphics[width=80mm]{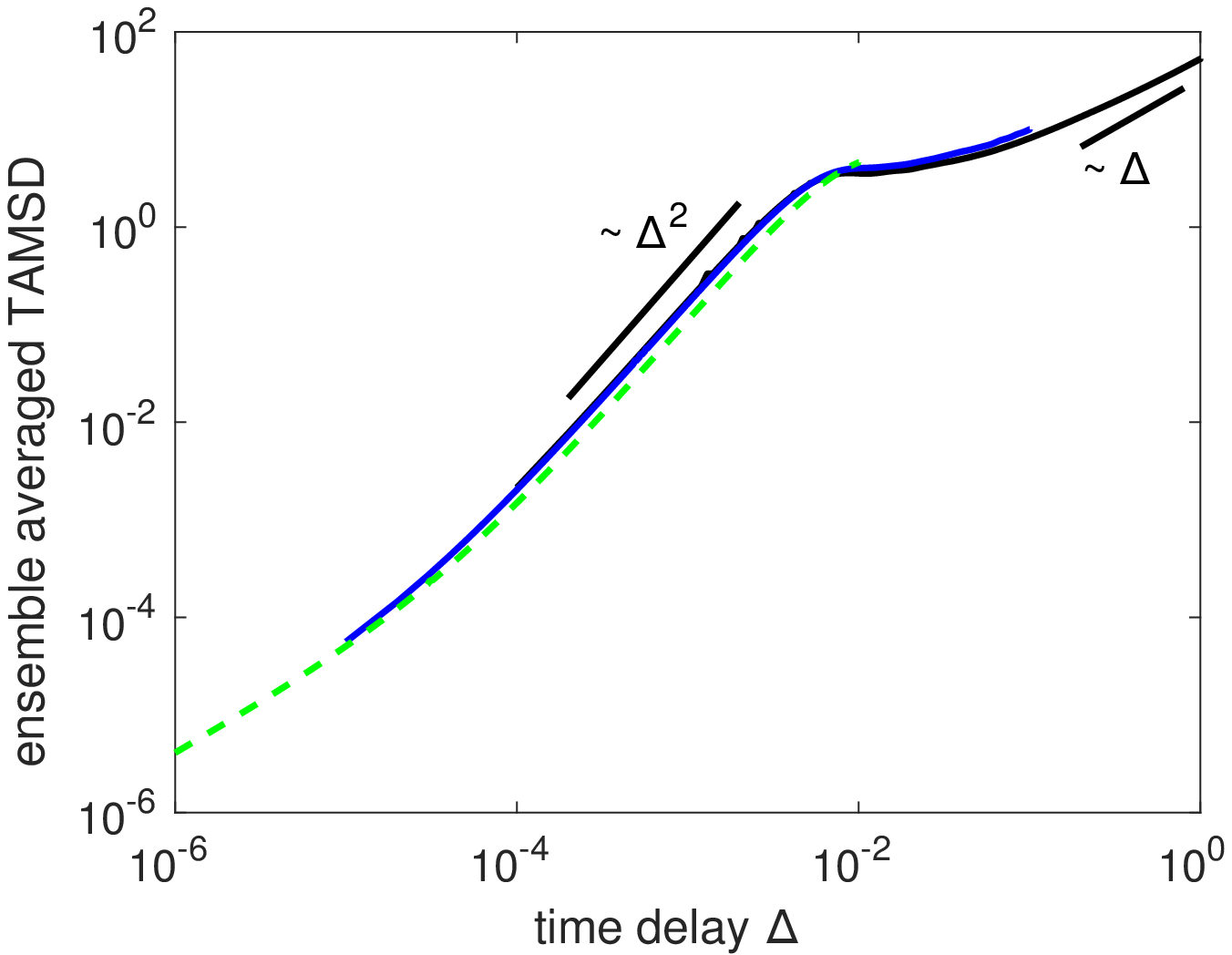}
\caption{$\langle\overline{R^2(\Delta, T)}\rangle$ according to Eq. (\ref{ATAMSD}) starting at the separatrix averaged over $10^2$ walks for the EL flow (left) and the YPP flow (right) for $T=10$ (black), $T=1$ (blue) and $T=0.1$ (green dashed).}
\end{figure*}

The intermediate time behavior of the aged MSD for $t_1 \ll t \ll t_2$ is the most interesting one. It is similar for both flows.
As one readily infers from Fig.~4
, the values of $\langle R^2(t,t_a) \rangle = \langle (\mathbf{r}(t_a+t) - \mathbf{r}(t_a))^2 \rangle$ in the time domain
above and for moderate $t_a \ll t_2$ are small compared to the squared cell size. Therefore 
this behavior is dominated by the complex dynamics within a single eddy, and shows oscillations, whose amplitude decays slowly with the increase of both 
$t$ and $t_a$. 
This kind of aging behavior is not observed in a comb model which does not show any internal dynamics within the trapped state. 
The oscillations exhibited by the aged MSD are due to the fact that the tracer position after aging time $t_a$ is not at the center but
at a finite distance from it. The further motion of the tracer approximately follows a closed streamline around the center of the cell, and
indeed the frequency of the oscillations corresponds to the angular velocity of such rotations, which (in our dimensionless units) follows from 
the solution of the deterministic part of Eq. (\ref{Lang1}) for $\mathbf{r}$ close to the center of the cell. These frequencies are $\omega = \mathrm{Pe}$ for the EL flow and 
$\omega = (4\pi)^{-1} \mathrm{Pe}$ 
for the YPP flow. Our simulations corroborate the findings for the corresponding time periods of the rotations for the examined $\mathrm{Pe}$ values. 
The decay of these oscillations is caused by the dephasing of the motion. In the course of time more and more particles move away from the center of the cell.
On the other hand, the circulation frequency depends on the distance from the center and decays to zero at its periphery. \\
\\
For the EL flow an approximate formula for the oscillatory behavior of the MSD can be derived in the following way: Sufficiently close to the center of the cell the streamlines are nearly circular, and the diffusion occurs basically in the radial direction, whereas angular distributions are practically uniform. Let the particles at time $t=-t_a$ be $\delta$-distributed at the center of the cell. At $t=0$ their probability density is then given by 
\[
 p(r) = \frac{1}{4\pi Dt_a}\exp\left(-\frac{r^2}{4Dt_a}\right)
\]
with $D=1$ in our units. Diffusion is assumed to be much slower than advection. Thus let us for the moment
``freeze'' the diffusion completely. We fix this distribution and assume that the tracers are uniformly advected along their respective circular streamlines. Then evolution of the MSD is governed by non-isochronicity of rotations. The larger the radius $r$, the longer the period $T(r)$. At time t, the instantaneous MSD for the infinitesimally thin ring of radius $r$ equals 
\[
 2r^2(1-\cos(\omega(r) t)) \times 2\pi r p(r)\, dr
\]
with $\omega(r)=2\pi/T(r)$. Thus the MSD equals $4\pi\int_{0}^{\infty} r^3 p(r)(1-\cos(\omega(r) t))\, dr$. When shifting the origin of the coordinates to the center of a cell, i.e. with $x=\pi/2+U/2$ and $y=\pi/2+V/2$, the equations $\dot{\mathbf{r}}=\mathrm{Pe}\; \mathrm{rot}(0,0,\Psi(\mathbf{r}))$ become $\ddot{U}+\mathrm{Pe}^2\sin(U)=0$ and $\ddot{V}+\mathrm{Pe}^2\sin(V)=0$. Solutions of these pendulum equations are elliptic functions. For the oscillation of variable $U$ with amplitude $U_0$, the period
equals $T(r)=\frac{4}{\mathrm{Pe}}K(1-\cos(U_0))$ with $K(m)$ being the complete elliptic integral of first kind, see Chap. 17 in \cite{Abramowitz}. For sufficiently small amplitudes the period obeys
\[
 T(r)=\frac{2\pi}{\mathrm{Pe}}+\frac{\pi U_0^2}{4\mathrm{Pe}}+\frac{19\pi U_0^4}{384\mathrm{Pe}}+\dots
\]
with $U_0=2r$. Considering only terms up to quadratic order in $r$, we end up with $\omega(r)=\frac{2\pi}{T(r)}=\mathrm{Pe}\left(1-\frac{r^2}{2}\right)$. Substituting into the expression for the MSD and performing the integration yields
\begin{eqnarray*}
&\langle R^2(t,t_a) \rangle = 8t_a + \displaystyle\frac{8t_a}{(1+4\mathrm{Pe}^2 t^2 t_a^2)^2} \\
&\times [(4\mathrm{Pe}^2 t^2 t_a^2-1)\cos(\mathrm{Pe}\, t)-4\mathrm{Pe}\,t t_a \sin(\mathrm{Pe}\,t)].
\end{eqnarray*}

Now we ``unfreeze'' the diffusion stopped at $t_a$, i.e. we replace $t_a$ by $t_a+t$. After the oscillations die out, the approximate expression becomes linear in time. Then the approximation does not hold any longer, since further terms in the expansion of the elliptic integral should be taken into account. Furthermore, the relevant streamlines of the EL flow are not circular anymore. However, for small times and small aging times, this approximation describes the MSD quite well, as we can see in Fig.~5
, where the oscillating part of the aged MSD from Fig.~4 
 is compared to our obtained formula. A similar derivation for the YPP flow leads to much more cumbersome expressions, since the streamlines are not
circular even in the vicinity of the center. \\
\\
The behavior of the time-averaged MSD, Eq. (\ref{TAMSD}), strongly depends on the total averaging time. Here we report the results for $T \gg t_2 = 1$.  
For this case the ensemble-averaged TAMSD for particles starting at the separatrix displays the behavior changing from ballistic at short times to diffusive at long times, as shown in Fig.~6 
 (black), very similar to the behavior observed for initially homogeneously flooded cells. This intermediate stage
disappears for low $\mathrm{Pe}$, when $t_1$ and $t_2$ get too close. The overall type of the behavior can be explained by the discussion of a single trajectory,
as shown in Fig.~1. 
 Building TAMSD with longer $T$ corresponds to averaging over all possible positions which the particle assumes during its
motion taken as initial position. The distribution of these positions, reduced to a single cell, is relatively homogeneous. 
Since for $T \gg t_2$ the system homogenizes, the dependence on $T$, as well as the dependence on initial conditions, disappear, and the distribution of 
TAMSD around its ensemble average is relatively narrow. The situation here is similar to the case of CTRW with truncated power-law WTD as
discussed in \cite{Akimoto1}. For $T$ of about order unity, see the green dashed and blue curves in Fig.~6
, the ensemble-averaged TAMSD is also already very similar to the MSD for the flooded case.
For $T \ll t_2$ the typical ergodicity breaking behavior should be present, but the dependence on the initial conditions and the complicated internal dynamics
within the trapped state make the situation much more involved than the one for pure CTRW, see \cite{Akimoto1,Akimoto2}.

\section{Conclusions}
\label{sect_conclusions}

We have considered the diffusion-advection problem in two simple cellular flow models that differ with respect to the boundary conditions imposed on the cell edges.
The models, often invoked as examples for subdiffusive tracer motion, were hardy investigated in detail.
We concentrate on the intermediate time range, in which the tracer's motion indeed may show subdiffusion. 
Extensive numerical simulations of the systems under different initial conditions show 
that the intermediate-time subdiffusion regime is only evident when the particles start at  
the border between different cells, i.e. at the separatrix, and is less pronounced or absent for other initial conditions,
e.g. when particles initially are injected in the cell center or are homogeneously distributed within the cell.
The complex motion of the particles within the single cell leads to  peculiar aging properties of the system in
this intermediate-time domain, and is mirrored also in the behavior of the time-averaged mean squared displacement for
single trajectories. Such behavior is not captured by classical models based on continuous time random walks that possess no dynamics in the trapped state. 

\section{Acknowledgements}

The work was financed within the project 3140-11899 funded by the German-Israeli Foundation for Scientific Research and Development (GIF). 
We are thankful to Prof. Eli Barkai, whose questions provoked the present work.

\end{document}